\documentclass[aps,prb,floats,twocolumn]{revtex4-1}
\usepackage{graphicx}
\usepackage{amssymb}
\usepackage{amsmath}
\usepackage{color}
\usepackage{float}
\graphicspath{{./FIG_FINALI/}}
\usepackage[caption=false]{subfig}

\begin{document}
\title{Charge transfer energies of benzene physisorbed on a graphene sheet from constrained density functional theory}
\author{Subhayan Roychoudhury}
\affiliation{School of Physics, AMBER and CRANN Institute, Trinity College Dublin, Dublin 2, Ireland}
\author{Carlo Motta}
\affiliation{School of Physics, AMBER and CRANN Institute, Trinity College Dublin, Dublin 2, Ireland}
\author{Stefano Sanvito}
\affiliation{School of Physics, AMBER and CRANN Institute, Trinity College Dublin, Dublin 2, Ireland}

\begin{abstract}
Constrained density functional theory (CDFT) is used to evaluate the energy level alignment of a benzene 
molecule as it approaches a graphene sheet. Within CDFT the problem is conveniently mapped onto 
evaluating total energy differences between different charge-separated states, and it does not consist 
in determining a quasi-particle spectrum. 
We demonstrate that the simple local density approximation provides a good description of the level aligmnent
along the entire binding curve, with excellent agreement to experiments at an infinite separation and to $GW$
calculations close to the bonding distance. The method also allows us to explore the effects due to the presence of
graphene structural defects and of multiple molecules. In general all our results can be reproduced by a
classical image charge model taking into account the finite dielectric constant of graphene.
\end{abstract}

\maketitle

\section{Introduction}

Organic molecular crystals, namely crystals composed of organic molecules held together by weak 
van der Waals forces, are emerging as excellent candidates for fabricating nanoscale devices. These
have potential application in electronics and optoelectronics in particular in areas such as solar energy
harvesting, surface photochemistry, organic electronics and spintronics~\cite{Dimitrakopoulos,Gadzuk,Repp,Koch,Sanvito}. 
A feature common to such class of devices is that they are composed from both an organic and inorganic 
component, where the first forms the active part of the device and the second provides the necessary
electrical contact to the external circuitry. Clearly the electronic structure of the interface between these 
two parts plays a crucial role in determining the final device performance and needs to be understood 
carefully. In particular it is important to determine how charge transfers between the organic and the
inorganic component and the energies at which the transfer takes place. This is a challenging task, 
especially in the single-molecule limit. Upon adsorption on a substrate, the electron addition 
and removal energies of a molecule change value from that of their gas phase counterparts. This is 
expected since, when the molecule is physisorbed on a polarisable substrate, the removal (addition) of 
an electron from (to) the molecule gives rise to a polarisation of the substrate. The image charge 
accumulated on the substrate in the vicinity of the molecule alters the addition or removal energy of 
charge carriers from the molecule.

A common way to calculate the addition and removal energies is to use a quasiparticle (QP) description. 
Within the QP picture, one ignores the effects of relaxation of molecular orbitals due to addition or removal 
of electrons and consequently takes the relative alignment of the metal Fermi level, $E_\mathrm{F}$, with either
the lowest unoccupied molecular orbital (LUMO) and highest occupied molecular orbital (HOMO) of a molecule 
as removal energy. This effectively corresponds to associate the electron affinity and the ionization potential
respectively to the LUMO and HOMO of the molecule. The adequacy of the QP description then depends
on the level of theory used to calculate the energy levels of the HOMO and LUMO.

If the theory of choice is density functional theory (DFT)~\cite{DFT}, then a number of observations should be
made. Firstly, it is important to note that except for the energy of the HOMO, which can be rigorously interpreted
as the negative of the ionization potential~\cite{Janak}, in general the Kohn-Sham orbitals cannot be associated
to QP energies. This is, however, commonly done in practice and often the Kohn-Sham QP levels provide a good 
approximation to the true removal energies, in particular in the case of metals. For molecules unfortunately the 
situation is less encouraging with the local and semi-local approximations of the exchange and correlation functional, 
namely the local density approximation (LDA) and the generalized gradient approximation (GGA), performing
rather poorly even for the HOMO level. Such situation is partially corrected by hybrid functionals~\cite{hybrid} or 
by functional explicitly including self-interaction corrections~\cite{SIC1,SIC2}, and extremely encouraging results
have been recently demonstrated for range separated functionals~\cite{Leeor1,Leeor2}.

The calculation of the energy levels alignment of a molecule in the proximity of a metal, however, presents additional
problems. In fact, the formation of the image charge, although it is essentially a classical electrostatic phenomenon, 
has a completely non-local nature. This means that unless a given functional is explicitly non-local it will in general fail
in capturing such effect. The most evident feature of such failure is that the position of the HOMO and LUMO 
changes very little when a molecule approaches a metallic surface~\cite{Thygesen}. Such failure is typical of the 
LDA and GGA, and both hybrid and self-interaction corrected functionals do not improve much the situation. 
A possible solution to the problem is that of using an explicit many-body approach to calculate the QP
spectrum. This is for instance the case of the GW approximation~\cite{Hybertsen}, which indeed is capable
of capturing the energy levels renormalization due to the image charge effect~\cite{Neaton}. The GW scheme,
however, is highly computationally demanding and can be applied only to rather small systems. This is not
the case for molecules on surfaces, where the typical simulation cells have to include several atomic layers
of the metal and they should be laterally large enough to contain the image charge in full. This, in addition to 
the GW necessity to compute a significant fraction of the empty states manifold, make the calculations demanding 
and it is often not simple even to establish whether convergence has been achieved.

In this paper we approach the problem of evaluating the charge transfer energies of an organic molecule 
physisorbed on an inorganic substrate with the help of a much more resource-efficient alternative, namely 
constrained density functional theory (CDFT)~\cite{Wu}. In CDFT, one transfers one electron from the molecule 
to the substrate (and vice versa) and calculates the difference in energy with respect to the locally charge
neutral configuration (no excess of charge either on the molecule or the substrate)~\cite{Souza}. As such
CDFT avoids the calculation of a QP spectrum, which is instead replaced by a series of total energy 
calculations for different charge distributions. This approach is free of any interpretative issues and benefits
from the fact that even at the LDA level the total energy is usually an accurate quantity. Finally, it is important
to remark that, for any given functional, CDFT is computationally no more demanding than a standard DFT 
calculation, so that both the LDA and the GGA allow one to treat large systems and to monitor systematically 
the approach to convergence.

Here we use the CDFT approach to study the adsorption of molecules on a 2-dimensional (2D) metal in 
various configurations. It must be noted that in contrast to a regular 3D metal, in a 2D one the image charge 
induced on the substrate is constrained within a one-atom thick sheet. This means that electron screening
is expected to be less efficient than in a standard 3D metal and the features of the image charge formation in 
general more complex. In particular we consider here the case of graphene, whose technological relevance
is largely established~\cite{FerrariReview}. Most importantly for our work, recently graphene has been used
as template layer for the growth of organic crystals~\cite{GrapheneTemp}. It is then quite important to
understand how such template layer affects the level alignment of the molecules with the metal.
%
%
As a model system we consider a simple benzene molecule adsorbed on a sheet of graphene. This has been studied
in the past~\cite{AlZahrani,Chakarova,Yong-HuiZhang}, so that a good description of the equilibrium distance and
the corresponding binding energy of the molecule in various configurations with respect to the graphene sheet are 
available. Furthermore, a $\rm{G_0W_0}$ study for some configurations exists~\cite{Despoja}, so that our calculated
QP gap can be benchmarked.

Our calculations show that the addition and removal energies decrease in absolute value as the molecule is 
brought closer to the graphene sheet. Such decrease can be described with a classical electrostatic model taking
into account the true graphene dielectric constant. As it will be discussed, a careful choice of the substrate unit 
cell is necessary to ensure the inclusion of the image charge, whose extension strongly depends on the 
molecule-substrate distance. We also reveal that the presence of defects in the graphene sheet, such as a 
Stone-Wales one, does not significantly alter the charge transfer energies. In realistic situations, \textit{e.g.} 
at the interface between a molecular crystal and an electrode, a molecule is surrounded by many others, 
which might alter the level alignment. We thus show calculations, where neighboring molecules are included 
above, below and in the same plane of the one under investigation. Interestingly, our results suggest that the 
charge transfer states are weakly affected by the presence of other molecules.

\section{Method}

In order to find the ground state energy of a system, Kohn-Sham DFT minimises a universal energy functional
\begin{equation}\label{equ1}
\begin{split}
E[\rho]=\sum_{\sigma}^{\alpha,\beta}\sum_{i}^{N_\sigma}\langle \phi_{i \sigma}|-\frac{1}{2}\nabla^{2}|\phi_{i \sigma}\rangle+\int d\mathbf{r}v_n(\mathbf{r})\rho(\mathbf{r})+J[\rho]\\
+E_\mathrm{xc}[\rho^{\alpha},\rho^{\beta}]\:,
\end{split}
\end{equation}
where $J$, $E_\mathrm{xc}$ and $v_n$ denote respectively the Hartee, exchange-correlation (XC) and external potential 
energies. The Kohn-Sham orbitals, $\phi_{i \sigma}$, for an electron with spin $\sigma$ define the non-interacting
kinetic energy $\langle \phi_{i \sigma}|-\frac{1}{2}\nabla^{2}|\phi_{i \sigma}\rangle$, while $N_{\sigma}$ is the 
total number of electrons with spin $\sigma$. The electron density, is then given by $\rho(\mathbf{r})=\sum_{\sigma}^{\alpha,\beta}\rho^{\sigma}=\sum_{\sigma}\sum_{i}^{N_{\sigma}}|\phi_{i\sigma}(\mathbf{r})|^2$.

In contrast to regular DFT, in CDFT one wants to find the ground state energy of the system subject to an additional 
constraint of the form
\begin{equation}\label{equ2}
\sum_{\sigma}\int w_\mathrm{c}^{\sigma}(\mathbf{r})\rho^{\sigma}(\mathbf{r})d\mathbf{r}=N_\mathrm{c}\:,
\end{equation}
where $w_\mathrm{c}^{\sigma}$ is a weighting function that describes the spatial extension of the constraining region 
and $N_\mathrm{c}$ is the number of electrons that one wants to confine in that region. In our case 
$w_\mathrm{c}^{\sigma}(\mathbf{r})$ is set to 1 inside a specified region and zero elsewhere. In order to 
minimise $E[\rho]$ subject to the constraint, we introduce a Lagrange multiplier $V_\mathrm{c}$ and define
the constrained functional \cite{Wu} 
\begin{equation}\label{equ3}
W[\rho,V_\mathrm{c}]=E[\rho]+V_\mathrm{c}\left(\sum_{\sigma}\int w_\mathrm{c}^{\sigma}(\mathbf{r})\rho^{\sigma}(\mathbf{r})d\mathbf{r}-N_\mathrm{c}\right)
\end{equation}

Now the task is that of finding the stationary point of $W[\rho,V_\mathrm{c}]$ under the normalization condition for the 
Kohn-Sham orbitals. This leads to a new set of Kohn-Sham equations
\begin{equation}\label{equ4}
\begin{split}
\left[-\frac{1}{2}\nabla^2+v_n(\mathbf{r})+\int \frac{\rho(\mathbf{r'})}{|\mathbf{r}-\mathbf{r'}|}d\mathbf{r'}+v_\mathrm{xc}^\sigma(\mathbf{r})+V_\mathrm{c}w_\mathrm{c}^{\sigma}(\mathbf{r})\right]\phi_{i\sigma}\\
=\epsilon_{i\sigma}\phi_{i\sigma}\:,
\end{split}
\end{equation}
where $v_\mathrm{xc}^\sigma(\mathbf{r})$ is the exchange and correlation potential. Equation~(\ref{equ4}) does
not compute $V_\mathrm{c}$, which remains a parameter. However, for each value of $V_\mathrm{c}$ it produces 
a unique set of orbitals corresponding to the minimum-energy density. In this sense we can treat $W[\rho,V_\mathrm{c}]$ 
as a functional of $V_\mathrm{c}$ only. It can be proved that $W[\rho,V_\mathrm{c}]$ has only one stationary point 
with respect to $V_\mathrm{c}$, where it is maximized~\cite{Wu}. Most importantly the stationary point satisfies the 
constraint. One can then design the following procedure to find the stationary point of $W[\rho,V_\mathrm{c}]$:
(i) start with an initial guess for $\phi_{i\sigma}$ and $V_\mathrm{c}$ and solve Eq.~(\ref{equ4}); 
(ii) update $V_\mathrm{c}$ until the constraint Eq.~(\ref{equ2}) is satisfied;
(iii) start over with the new $V_\mathrm{c}$ and a new set of $\phi$s.

Here we use CDFT to calculate the charge transfer energy between a benzene molecule and a graphene sheet. 
For any given molecule-to-substrate distance, $d$, we need to perform three different calculations:
\begin{enumerate}
\item a regular DFT calculation in order to determine the ground state total energy $E_{0}(d)$ and the amount of 
charge on each subsystem (\textit{i.e.} on the molecule and on the graphene sheet)
\item a CDFT calculation with the constraint that the graphene sheet contains one extra electron and the molecule 
contains one hole. This gives the energy $E_{+}(d)$.
\item a CDFT calculation with the constraint that the graphene sheet contains one extra hole and the molecule one 
extra electron. This gives the energy $E_{-}(d)$.
\end{enumerate}
The charge transfer energy for removing an electron from the molecule and placing it on the graphene sheet 
is then $E_{\rm CT}^{+}(d)=E_{+}(d)-E_{0}(d)$. Similarly, that for the transfer of an electron from the graphene 
sheet to the molecule is $E_{\rm CT}^{-}(d)=E_{0}(d)-E_{-}(d)$. Since in each run the cell remains charge
neutral, there is no need here to apply any additional corrections. However, we have to keep in mind that this 
method is best used when the two subsystems are well separated so that the amount of charge localized on 
each subsystem is a well defined quantity.

In our calculations we use the CDFT implementation~\cite{Souza} for the popular DFT package 
SIESTA\cite{Soler}, which adopts a basis set formed by a linear combination of atomic orbitals (LCAO). 
The constrain is introduced in the form of a projection over a specified set of basis orbitals and 
in particular use the Lowdin projection scheme. Throughout this work we adopt double-zeta polarized 
basis set with an energy cutoff of 0.02~Ry. The calculations are done with norm-conserving 
pseudopotential and the LDA is the exchange-correlation functional of choice. 
A mesh cutoff of 300~Ry has been used for the real-space grid. We impose periodic boundary conditions with 
different cell-sizes and the $k$-space grid is varied in accordance with the size of the unit cell. For instance, an 
in plane 5$\times$5 $k$-grid has been used for a 13$\times$13 graphene supercell.

\section*{Result And Discussion}
\subsection*{Single Benzene on pristine Graphene}
\begin{figure}
        \centering
\subfloat[]{\label{fig:hollow}%
  \fbox{\includegraphics[width=0.21\textwidth]{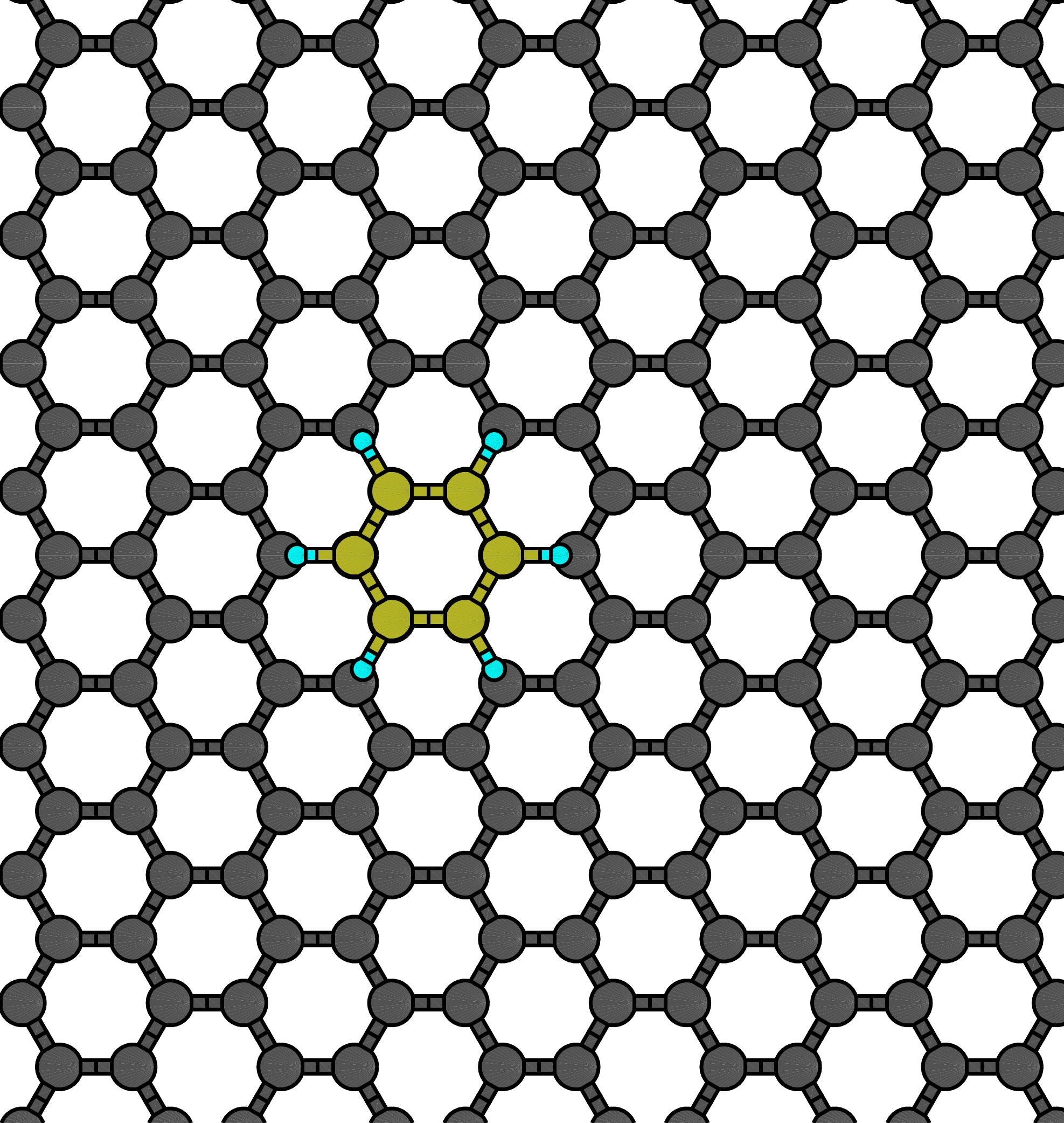}}%
}
\subfloat[]{\label{fig:stack}%
  \fbox{\includegraphics[width=0.21\textwidth]{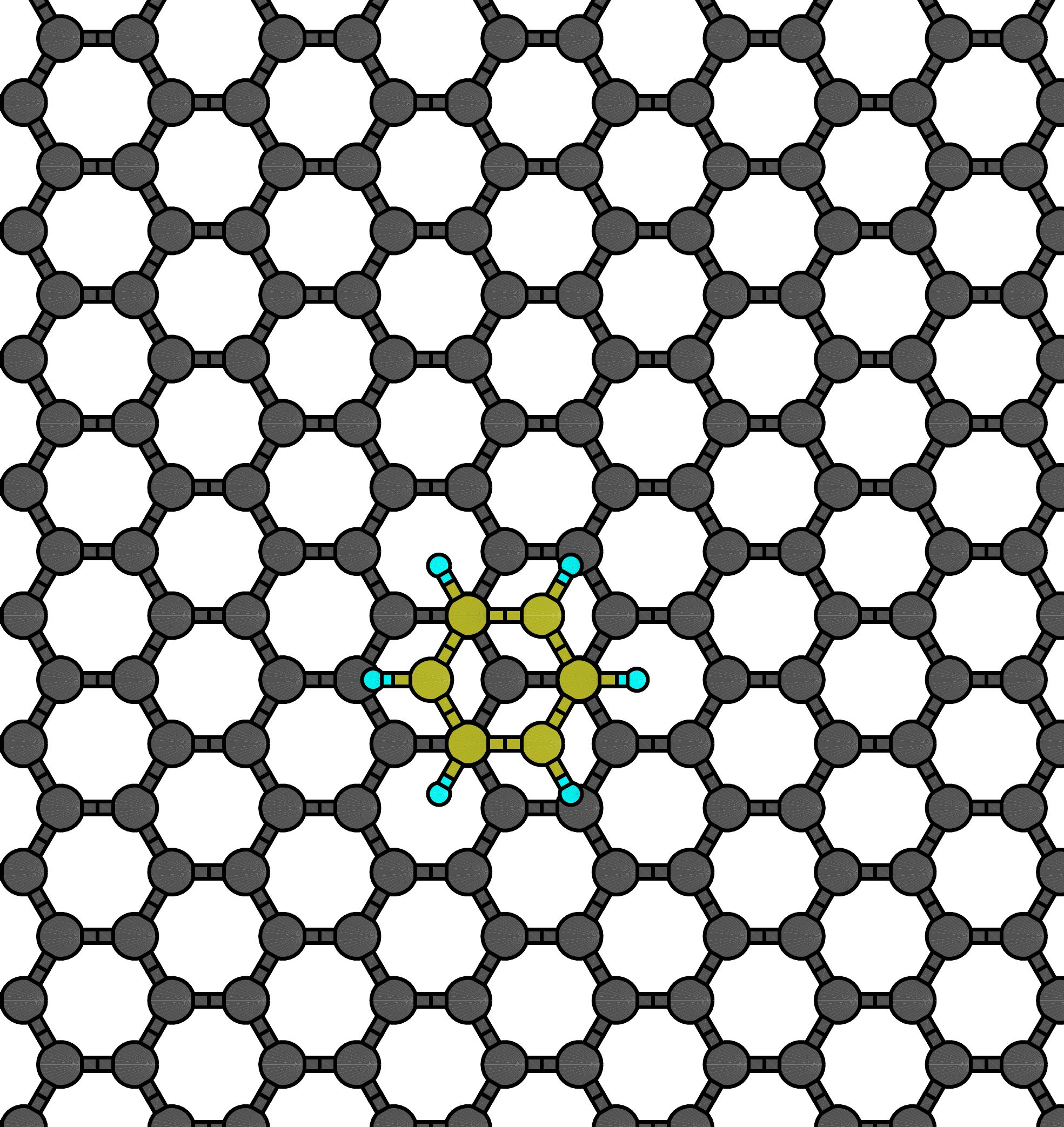}}%
}     

\subfloat[]{\label{fig:top_sw}%
  \fbox{\includegraphics[width=0.224\textwidth]{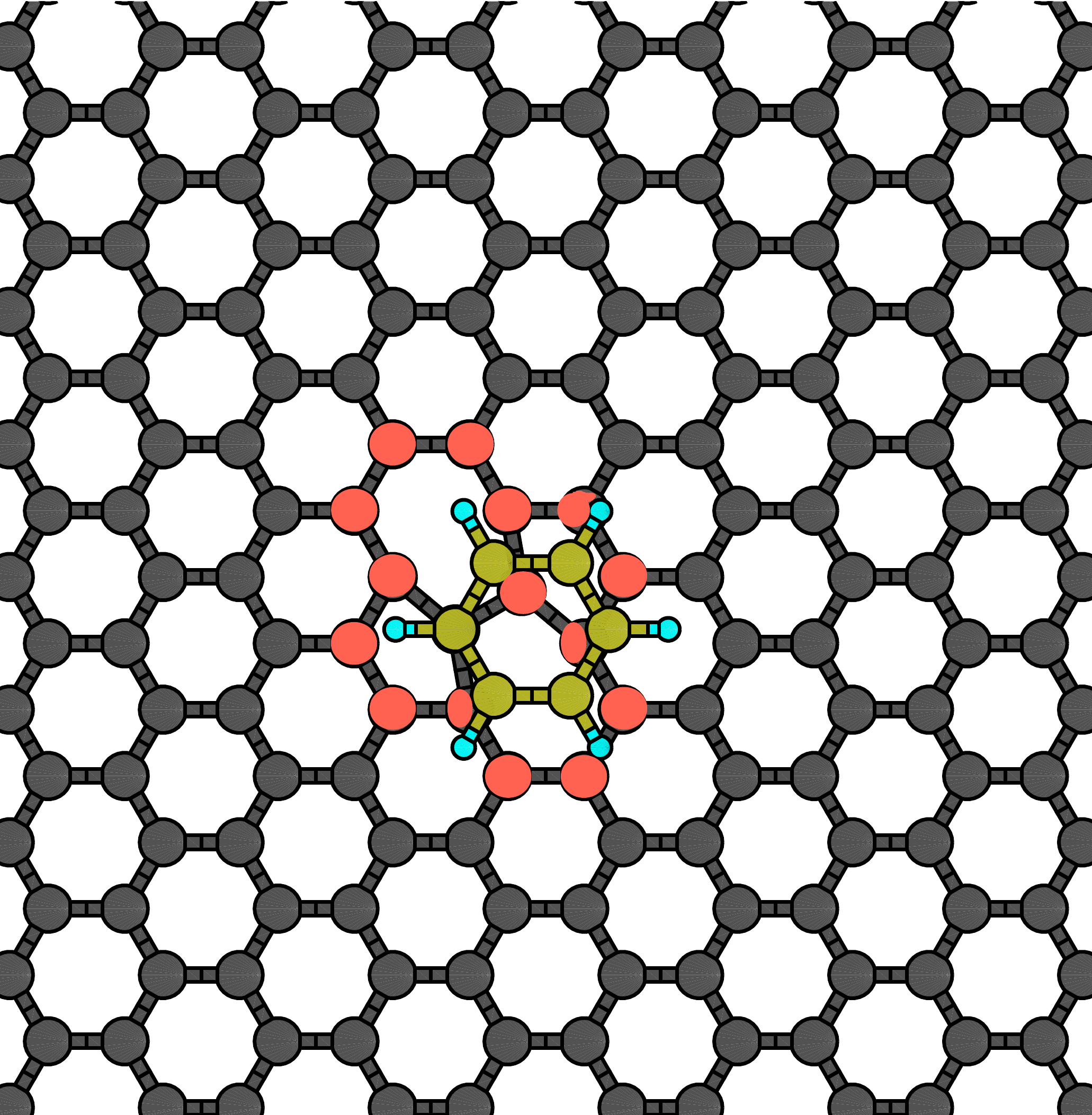}}%
}
\subfloat[]{\label{fig:away_sw}%
  \fbox{\includegraphics[width=0.21\textwidth]{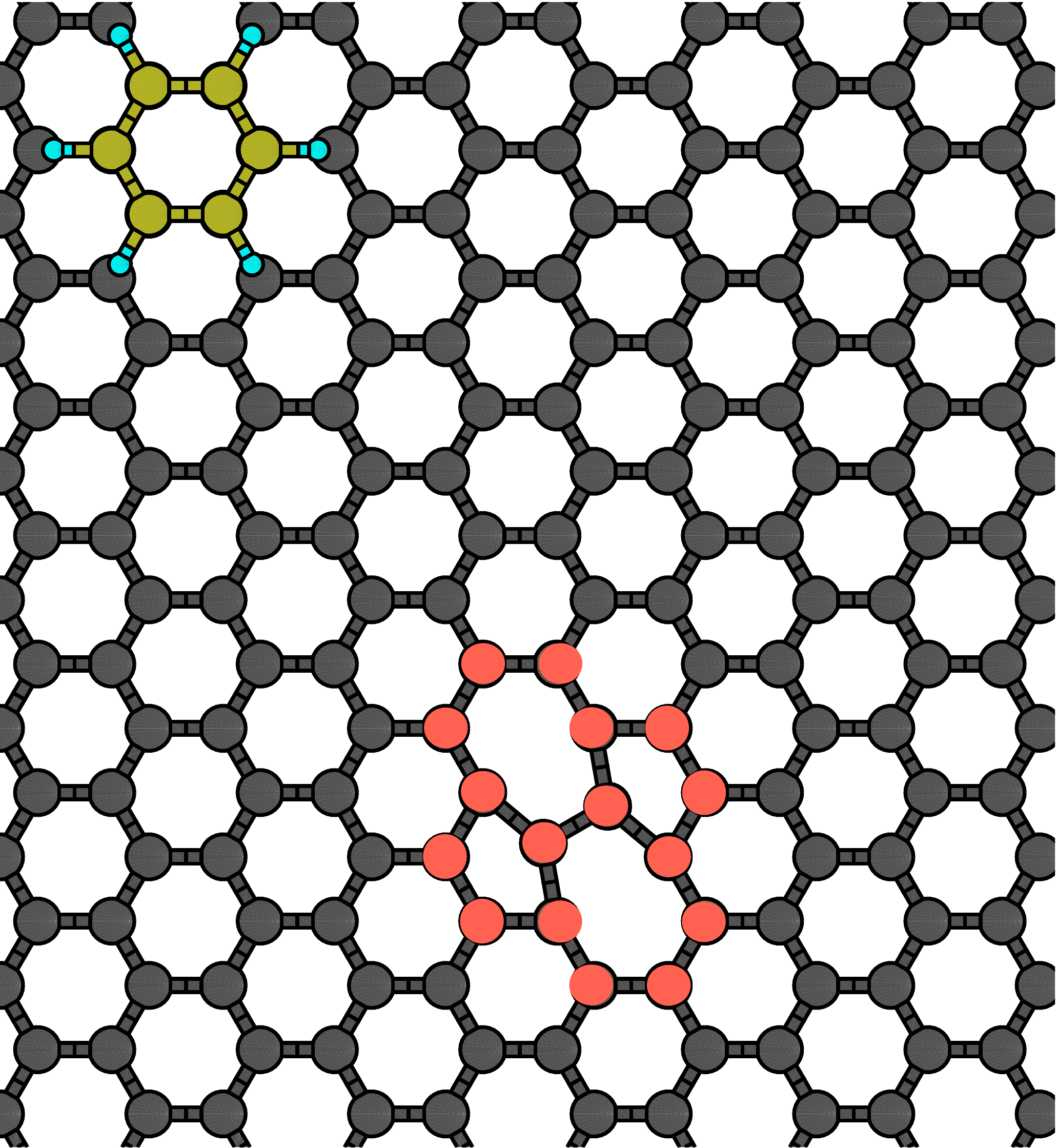}}%
}
        \caption{Top view of the (a) H, (b) S, (c) $\rm{H_{sw-T}}$ and (d) $\rm{H_{sw-A}}$ configurations. Green 
        spheres are carbon atoms belonging to benzene ring, while grey shperes are those of the graphene sheet. In the
        graphene sheet with a SW defect, the carbon atoms in all rings affected by the deformation are marked in red.}\label{fig:figure1}
\end{figure}

We begin this section with a discussion on the equilibrium distance for a benzene molecule adsorbed on graphene. 
This is obtained by simply minimizing the total energy difference 
$\Delta E_{\rm int}=E_{\rm Benz+Gr}-(E_{\rm Benz}+E_{\rm Gr})$, where $E_{\rm Benz+Gr}$ is the total energy
for the cell containing benzene on graphene, while $E_{\rm Benz}$ ($E_{\rm Gr}$) is the total energy of the 
same cell when only the benzene (graphene) is present. This minimization is performed for two different 
orientations of the benzene molecule with respect to the graphene sheet: the \textit{hollow} (H) configuration, 
in which all the carbon atoms of the benzene ring are placed exactly above the carbon atoms of graphene, 
and the \textit{stack} (S) configuration, in which alternate carbon atoms of the benzene molecule are placed 
directly above carbon atoms of the graphene sheet [see Fig.~\ref{fig:figure1}(a,b)]. For the H configuration
we find an equilibrium distance of 3.4~\AA, while for the S one this becomes 3.25~\AA. These results are in 
fair agreement with another LDA theoretical study~\cite{Yong-HuiZhang} (predicting 3.4~{\AA} and 3.17~{\AA} 
respectively for for the H and S orientations). Note that a more precise evaluation of such distances requires
the use of van der Waals corrected functionals. This exercise, however, is outside the scope of our work and
here we just wish to establish that the equilibrium distance is large enough for our constrain to remain well
defined. It can also be noted that the equilibrium distance of 3.6~{\AA} obtained with a vdW-DF study~\cite{Chakarova} is not very different from our LDA result.

\begin{figure}
\includegraphics[width=0.45\textwidth]{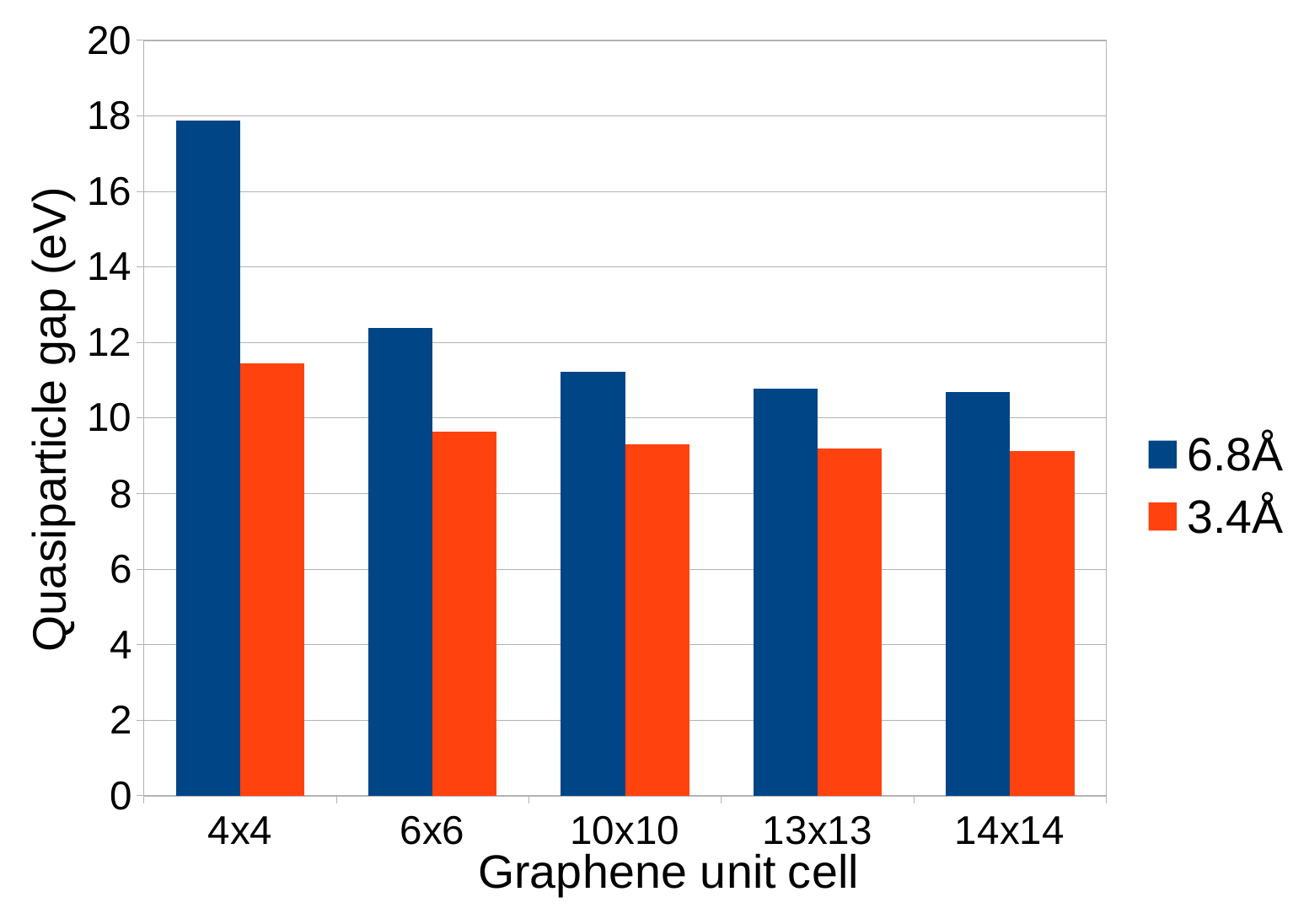}
\caption{Charge transfer energy gap, $E^+_{\rm CT}-E^-_{\rm CT}$, for different unit cell sizes of graphene sheet. 
The results are presented for two different molecule-to-graphene distances: 3.4~\AA\ and 6.8~\AA.}\label{fig:figure2}
\end{figure}
We then study the dependence of the charge transfer energies on the size of the graphene unit cell used. 
This is achieved by looking at the charge transfer gap, $E_{\rm CT}^{+}(d)-E_{\rm CT}^{-}(d)$, as a function 
of the unit cell size at various molecule-to-graphene distances (see Fig.~\ref{fig:figure2}).
When the molecule is very close to the graphene sheet, after transferring an electron, the image charge 
is strongly attracted by the oppositely charged molecule and thereby remains highly localized. However, 
as the molecule moves away from the substrate, the attraction reduces since the Coulomb potential 
decays with distance, resulting in a delocalization of the image charge. This will eventually spread uniformly 
all over the graphene sheet in the limit of an infinite distance. If the unit cell is too small, the image charge 
will be artificially over-confined, resulting in an overestimation of $E_-(d)$ and $E_+(d))$ and, as a 
consequence, of the charge transfer energies. This effect can be clearly seen in Fig.~\ref{fig:figure2},
where we display the variation of the charge transfer energies as a function of the cell size. Clearly, for the 
shorter distance (3.4~\AA\, corresponding to the average equilibrium distance), the energy gap converges
for supercells of about 10$\times$10 (10$\times$10 graphene primitive cells). At the larger distance of 
6.8~\AA\ the same convergence is achieved for a 13$\times$13 supercell. 
 
\begin{figure}
        \centering
\subfloat[]{\label{fig:f3_1}%
  \includegraphics[width=0.35\textwidth]{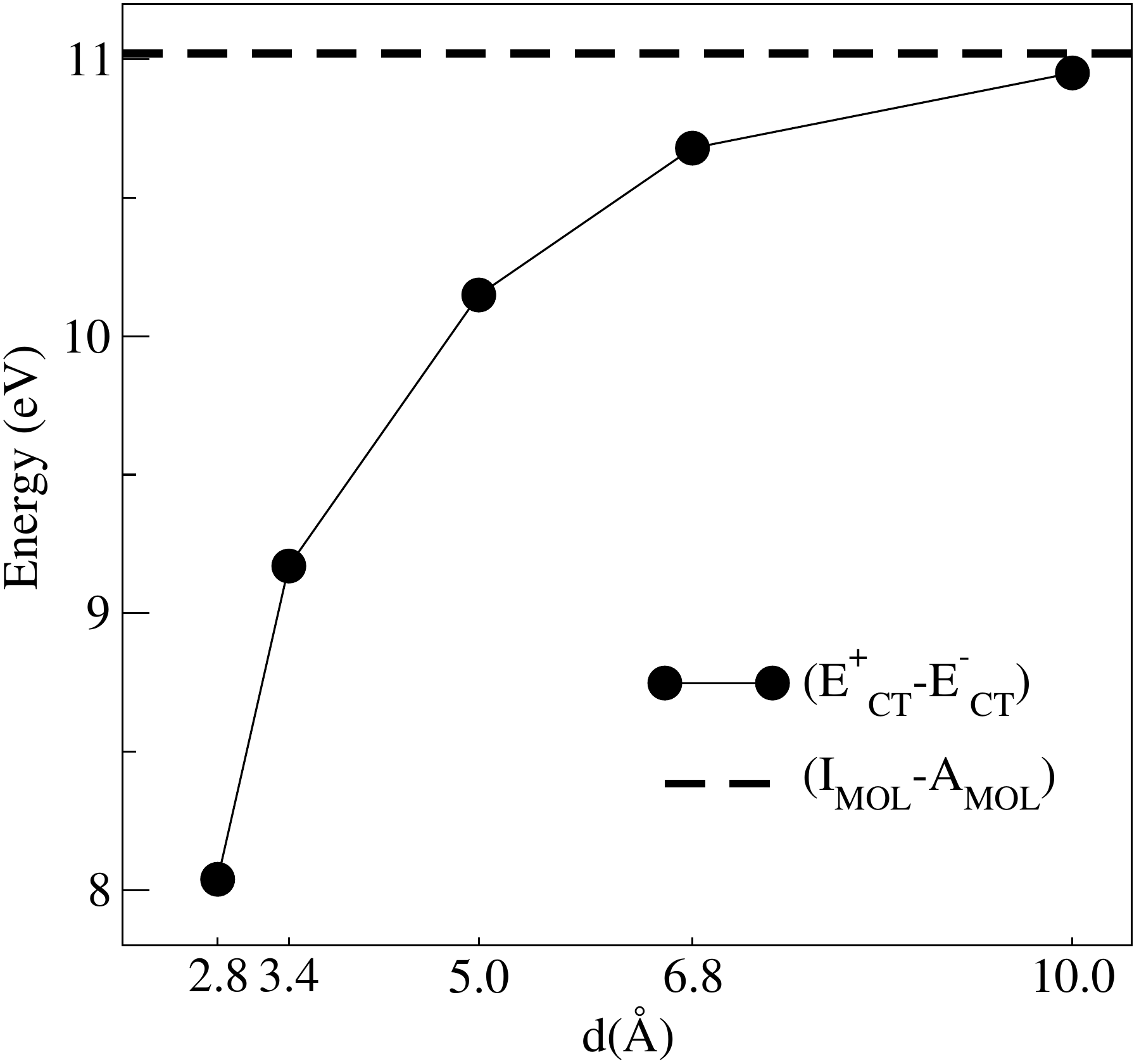}%
}
       
        ~ 
\subfloat[]{\label{fig:f3_2}%
  \fbox{\includegraphics[width=0.22\textwidth]{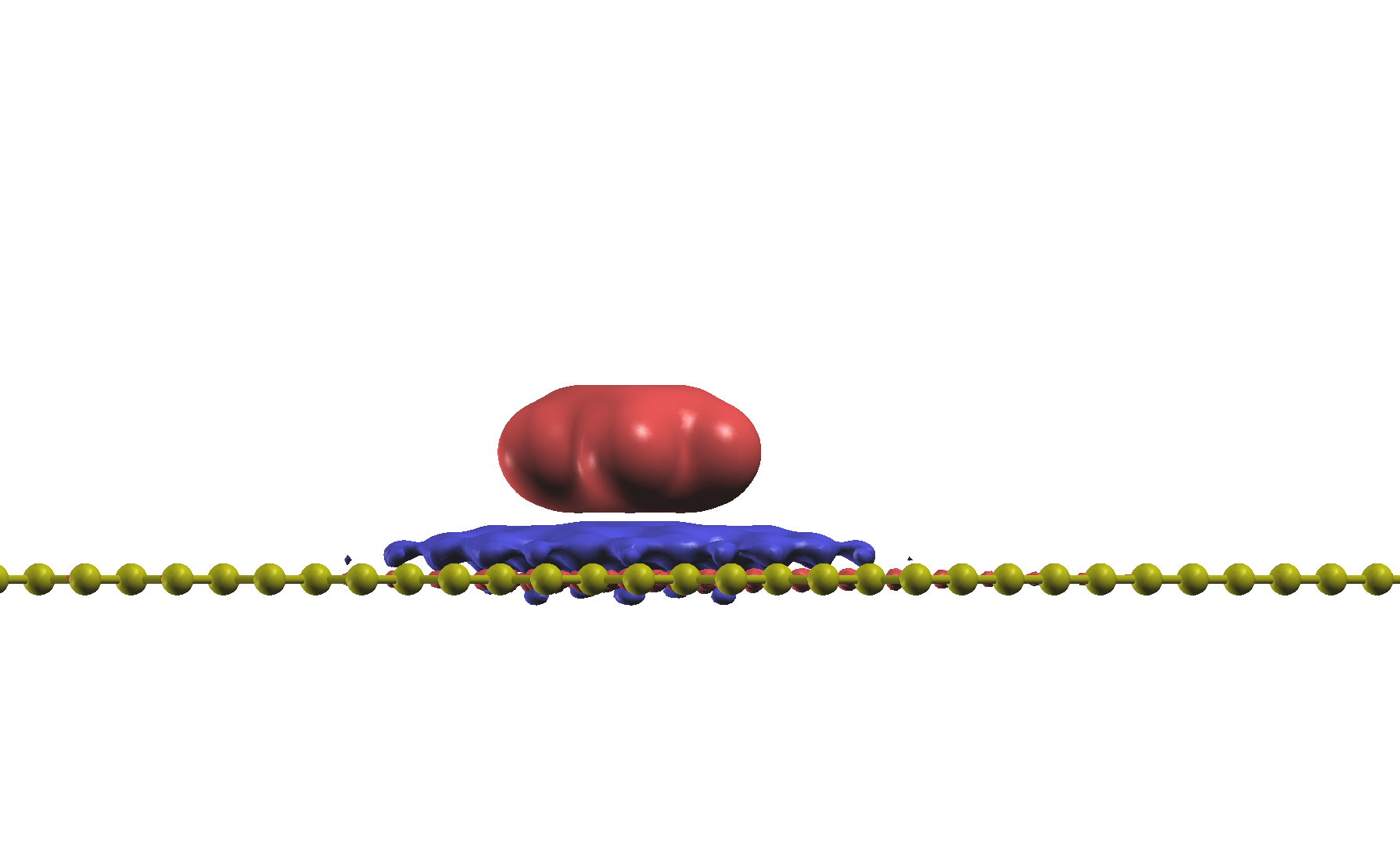}}%
}
\subfloat[]{\label{fig:f3_3}%
  \fbox{\includegraphics[width=0.22\textwidth]{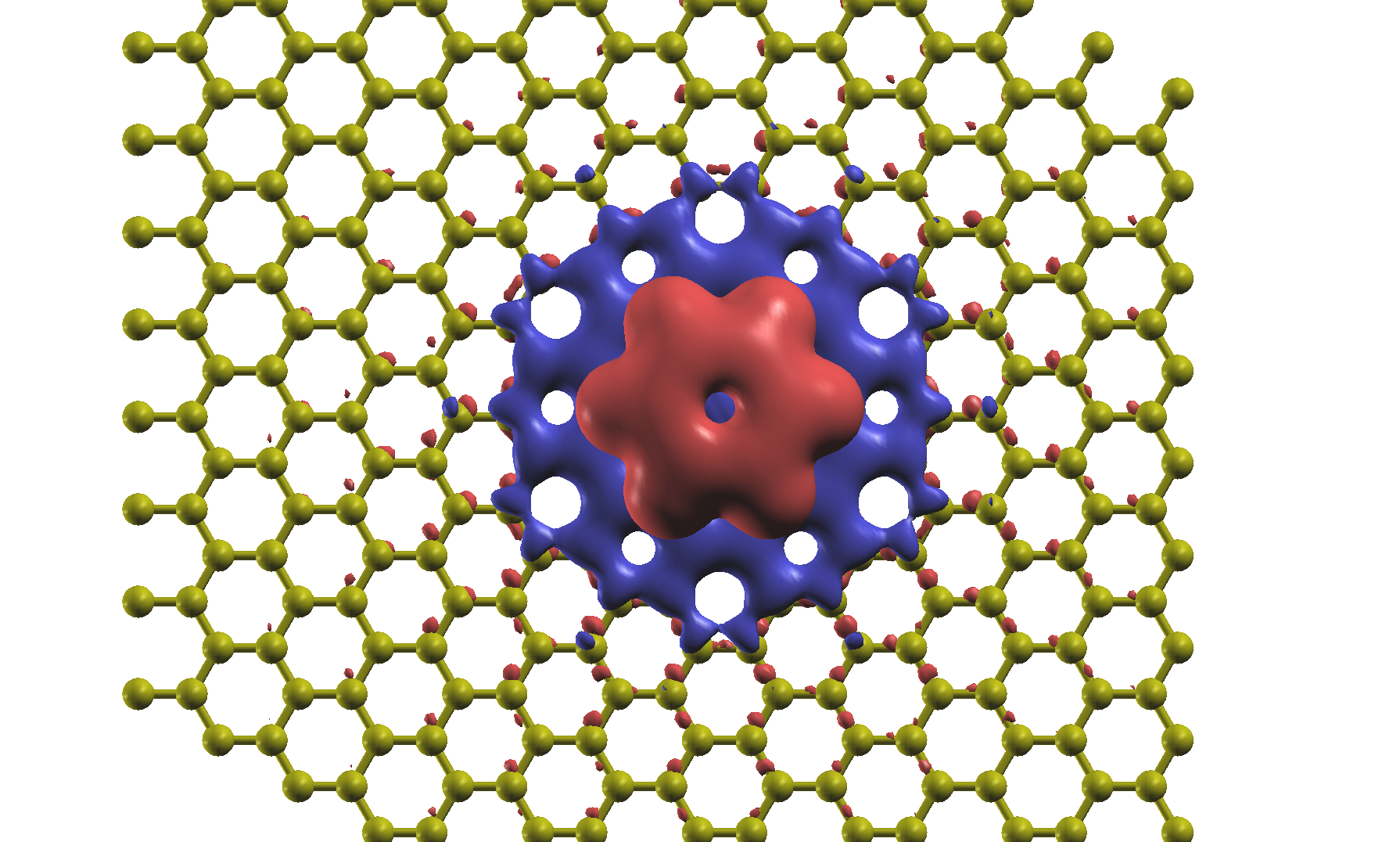}}%
}

\subfloat[]{\label{fig:f3_4}%
  \fbox{\includegraphics[width=0.22\textwidth]{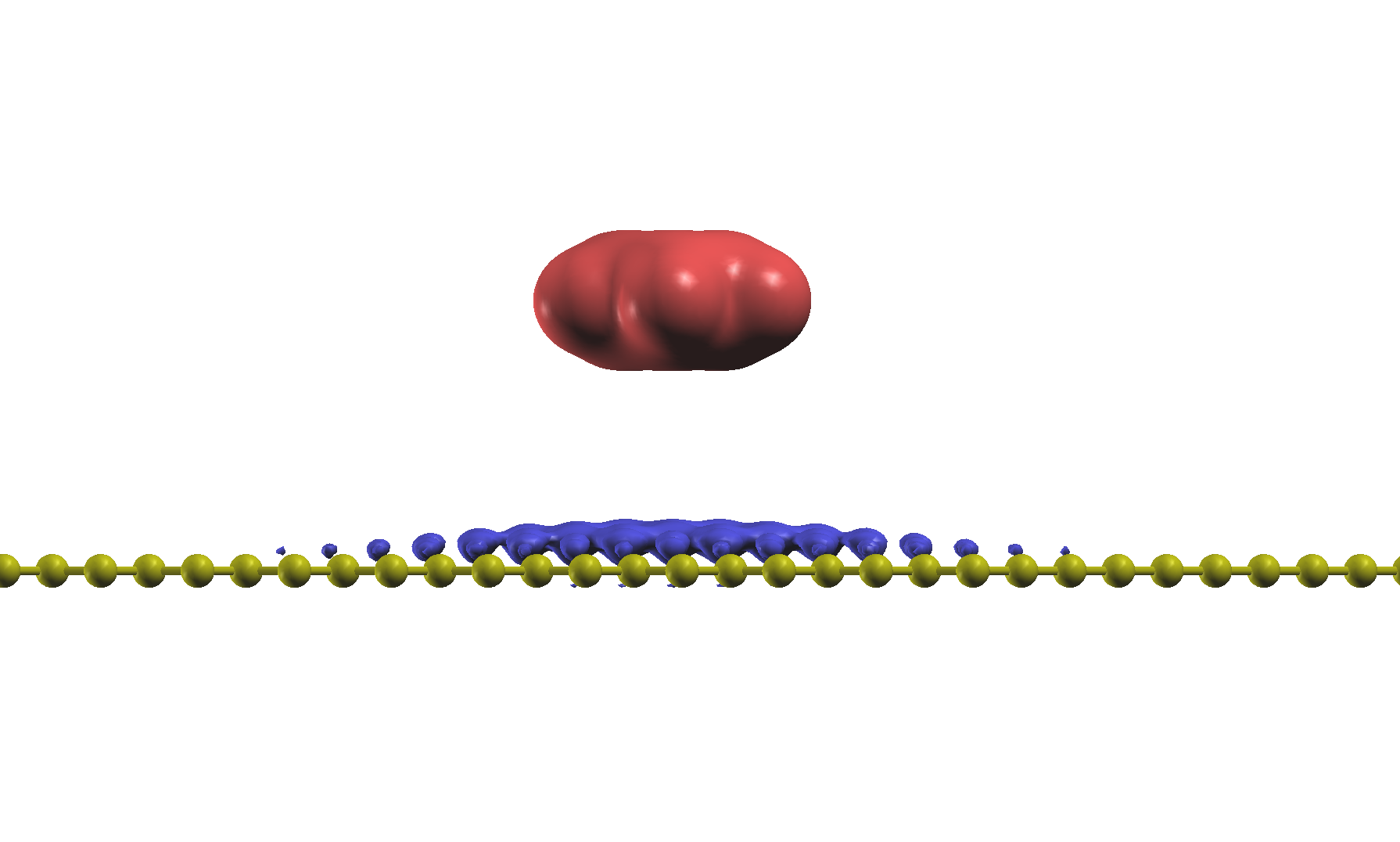}}%
}
\subfloat[]{\label{fig:f3_5}%
  \fbox{\includegraphics[width=0.22\textwidth]{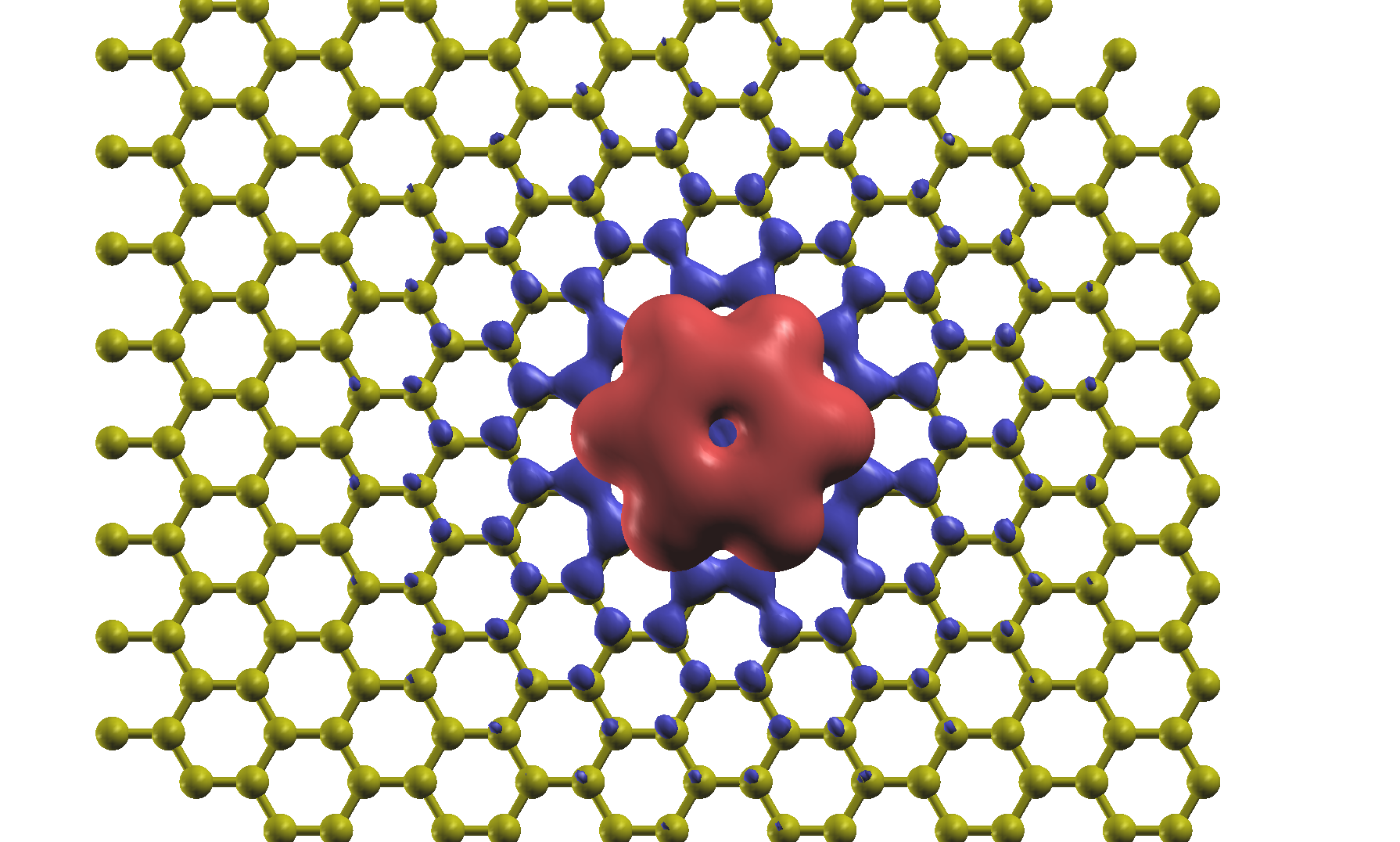}}%
}
\caption{Renormalization of the charge transfer gap as a function of the distance between the molecule
and the graphene sheet (a). The dashed line denotes the difference between the IP and the EA calculated 
with the $\Delta$SCF method. Panels (b) and (c) show the excess charge in different parts of the system 
after the transfer of an electron from the molecule to the sheet for $d=3.4$~\AA. Panels (d) and (e) show 
the same plot for $d=6.8$~\AA. In both cases, red and blue denote positive and negative net charge, 
respectively. In practice the charge distribution over the graphene plane corresponds to the charge
distribution of the image charge.}\label{fig:figure3}
\end{figure}

Next we compute the charge transfer energies as a function of the distance between the sheet and the 
molecule. In order to compare our results with the gap expected in the limit of an infinite distance, we 
need to evaluate first the ionization potential, $I_{\rm MOL}$, and the electron affinity, $A_{\rm MOL}$,
of the isolated benzene molecule. This is also obtained in terms of total energy differences between the
neutral and the positively and negatively charged molecule, namely with the $\Delta$SCF method. This
returns a quasiparticle energy gap, $I_{\rm MOL}-A_{\rm MOL}$, of 11.02~eV, in good agreement 
(within 4.5\%) with the experimental value~\cite{Clare,Burrow}. Likewise we also determine the Fermi 
level ($W_\mathrm{F}$) of graphene, which is found to be ~4.45 eV. 

In Fig.~\ref{fig:figure3}(a) we show the change in the charge transfer energy gap with the distance 
of the benzene from the graphene sheet for the H configuration. As expected, when the molecule is close 
to the surface, there is a considerably large attraction between the image charge and the opposite charge 
excess on the molecule, resulting in an additional stabilization of the system and a reduction in magnitude 
of $E_+(d)$ and $E_-(d)$. Hence, in such case the charge transfer energies have a reduced magnitude
and the charge transfer gap is smaller than that in the gas phase. 
Then, as the molecule moves away from the graphene sheet, the charge transfer energies increase and so
does the charge transfer energy gap until it eventually reaches the value corresponding to the HOMO-LUMO 
gap of the isolated molecule in the limit of an infinite distance. In Figs.~\ref{fig:figure3}(b), (c) ,(d) 
and (e) we show the excess charge-density, $\Delta \rho$, in different parts of the system after transferring 
one electron for two different molecule-to-graphene distances. The excess charge-density $\Delta \rho$ is 
defined as $\rho_{\rm CDFT}-\rho_{\rm DFT}$, where $\rho_{\rm DFT}$ and $\rho_{\rm CDFT}$ are 
respectively the charge densities of the system before and after the charge transfer. Thus the portion of 
$\Delta \rho$ localized on the graphene sheet effectively corresponds to the image charge profile.
Clearly, due to the stronger Coulomb attraction, the image charge is more localized for $d=3.4$~{\AA} 
than for $d=6.8$~\AA. At equilibrium for the S configuration, $d=3.25$~\AA, the charge 
transfer energy gap is calculated to be 8.91~eV, which is in good agreement (within 4\%) with the gap 
obtained by $G_0W_0$~\cite{Despoja}. In Table~\ref{table:configuration}, for the purpose of comparison, 
we have listed the charge transfer energies and charge transfer gaps for two different heights, 3.4~\AA\ 
and 6.8~\AA, and in different configurations. The most notable feature is that for the case of a pristine
graphene substrate the specific absorption site plays little role in determining the charge transfer levels
alignment.

\subsection*{Benzene on graphene with Stone-Wales defect}
In general actual graphene samples always display lattice imperfections \cite{Banhart}. In order to determine 
the effect of such structural defects on the CT energies, we consider a reference system where a Stone-Wales (SW)
defect (in which a single C-C bond is rotated by 90$^\circ$) is present in the graphene sheet. We have then 
calculated $E_{\rm CT}^{\pm}$ for two different positions of the molecule with respect to the defect on the 
sheet, namely the $\rm{H_{sw-T}}$ position, in which the molecule is placed right above the defect and the 
$\rm{H_{sw-A}}$ position, in which it is placed above the sheet far from the defect (see Fig.~\ref{fig:figure1}).
 
\begin{figure}
 \centering
  \includegraphics[width=0.44\textwidth]{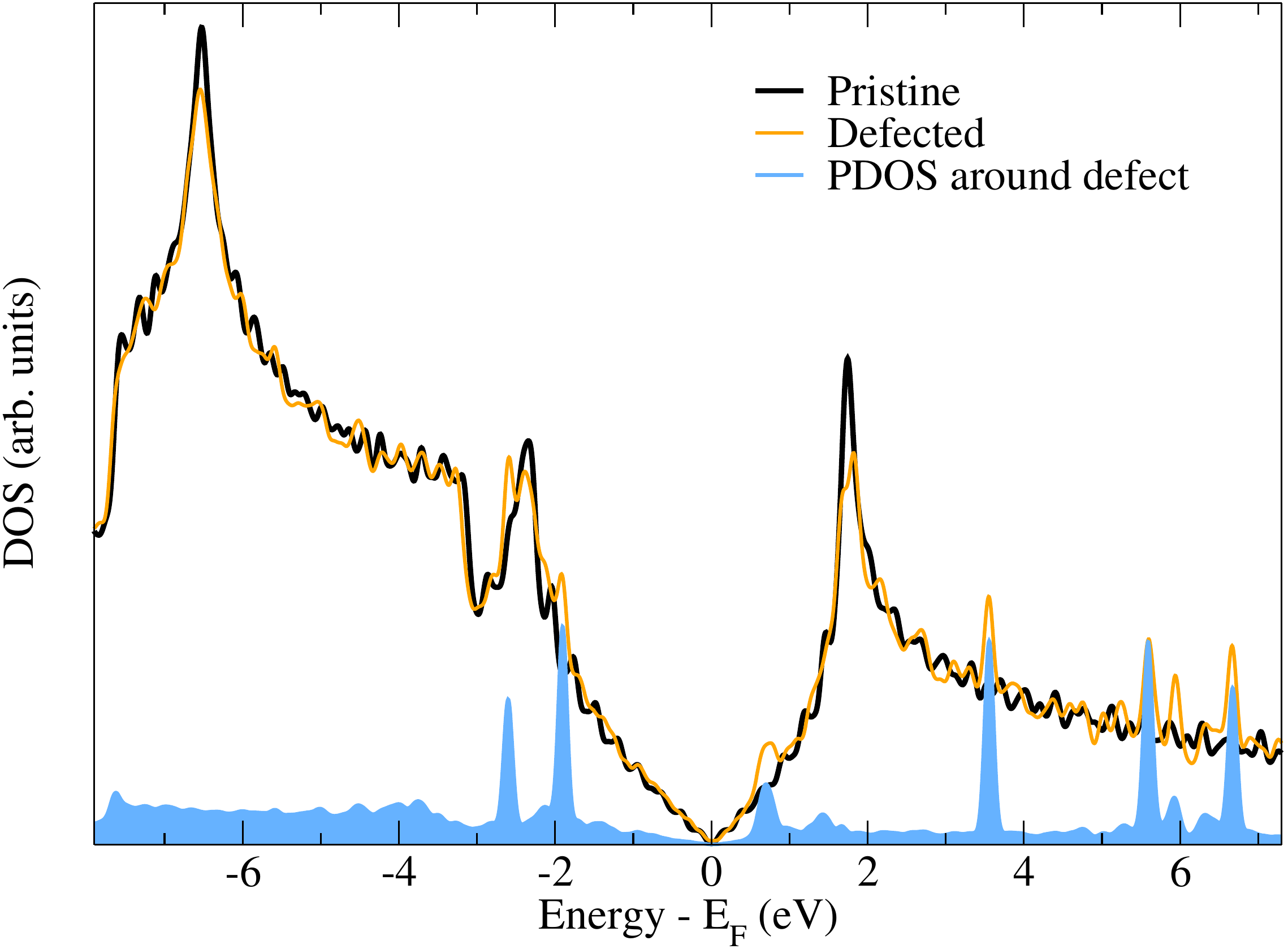}
  \captionsetup{justification=justified, singlelinecheck=false}
  \caption{DOS of pristine graphene (black line), DOS of a graphene sheet with one SW defect (orange line) and PDOS for the atoms 
  adjacent to the SW defect (blue shade). The latter has been multiplied by 3 for better visibility. These calculations are performed with a supercell of 200 
  atoms containing one SW defect.}
  \label{figure4}
 \end{figure}

Our findings are listed in Tab.~\ref{table:configuration}, where we report the charge transfer energies for both 
the configurations, assuming the molecule is kept at the same distance from the graphene sheet. From the table 
it is evident that the structural change in graphene due to presence of such defect does not alter the charge
transfer energies of the molecule. This is because the image charge distribution on graphene is little 
affected by presence of the SW defect. In addition, the density of states (DOS) of graphene remains almost 
completely unchanged near its Fermi energy after introducing such defect as can be seen in 
Fig.~\ref{figure4}, which shows that the partial density of states (PDOS) of the atoms forming the SW 
defect has no significant presence near the Fermi level. Thus, after the charge transfer, the electron added 
to (or removed from) the graphene sheet has the same energy that it would have in the absence of the defect,
i.e. it is subtracted (added) from a region of the DOS where there is no contribution from the SW defect. 
In this context, it is noteworthy that a $G_0W_0$ study \cite{Despoja} has concluded that altering the structure 
of pristine graphene by introducing dopant (which raises the Fermi level of graphene by 1~eV) also has minor 
effect on the QP gap of benzene, reducing it by less than 3\%. 

   \begin{table}
   \centering
    \begin{tabular}{ccccc}
    \hline \hline \vspace{0.5mm}
    Configuration & $d$ ({\AA}) & $E_{\rm CT}^+$ (eV) & $E_{\rm CT}^-$ (eV) & $\Delta E_{\rm CT}$ (eV) \\ \hline 
    {H} & 3.4 & 4.32 & -4.81 & 9.13 \\ 
    {H} & 6.8 & 4.95 & -5.72 & 10.67 \\ 
    $\rm{H_{sw-T}}$ & 3.4 & 4.30 & -4.78 & 9.08 \\ 
    $\rm{H_{sw-A}}$& 3.4 &4.34& -4.83 & 9.17 \\ 
    {S} & 3.4 & 4.33 & -4.81 & 9.14 \\ \hline 
    $\rm{H_{M1}}$ & 3.4 & 4.12 & -4.76 & 8.88 \\
    $\rm{H_{M2}}$ & 6.8 & 4.73 & -5.64 & 10.37 \\
    $\rm{H_L}$ & 3.4 & 4.27 & -4.62 & 8.89\\ \hline \hline
    \end{tabular}
     \captionsetup{justification=justified, singlelinecheck=false}
     \caption{$E_{\rm CT}^+$, $E_{\rm CT}^-$ and $\Delta E_{\rm CT}=E_{\rm CT}^+-E_{\rm CT}^-$ 
     for various configurations of a benzene molecule on pristine and defective graphene. H and S denote 
     adsorption of benzene on graphene in the \textit{hollow} and \textit{stack} configuration, respectively. 
     $\rm{H_{sw-T}}$ and $\rm{H_{sw-A}}$ correspond to adsorption on graphene with SW defect, with the 
     former corresponding to adsorption exactly on top of the defect and the latter corresponding to adsorption 
     away from the site of the defect. The configurations $\rm{H_{M1}}$ and $\rm{H_{M2}}$ both correspond 
     to adsorption of two benzene molecules in \textit{hollow} configuration- one at height 3.4~\AA\ and another 
     at a height 6.8~{\AA}. While in $\rm{H_{M1}}$, the CT is calculated for the lower molecule, in $\rm{H_{M2}}$, 
     the CT is calculated for the upper one. Finally $\rm{H_L}$ represents the case in which we have a layer of 
     non-overlapping benzenes adsorbed on graphene and one is interested in calculating the CT energy for 
     one of them, which is placed in the \textit{hollow} configuration.}
       \label{table:configuration}
    \end{table}

\subsection*{Multiple benzene molecules adsorbed on graphene}
\begin{figure}
        \centering

        ~ 
\subfloat[]{\label{fig:f5_1}%
  \fbox{\includegraphics[width=0.22\textwidth]{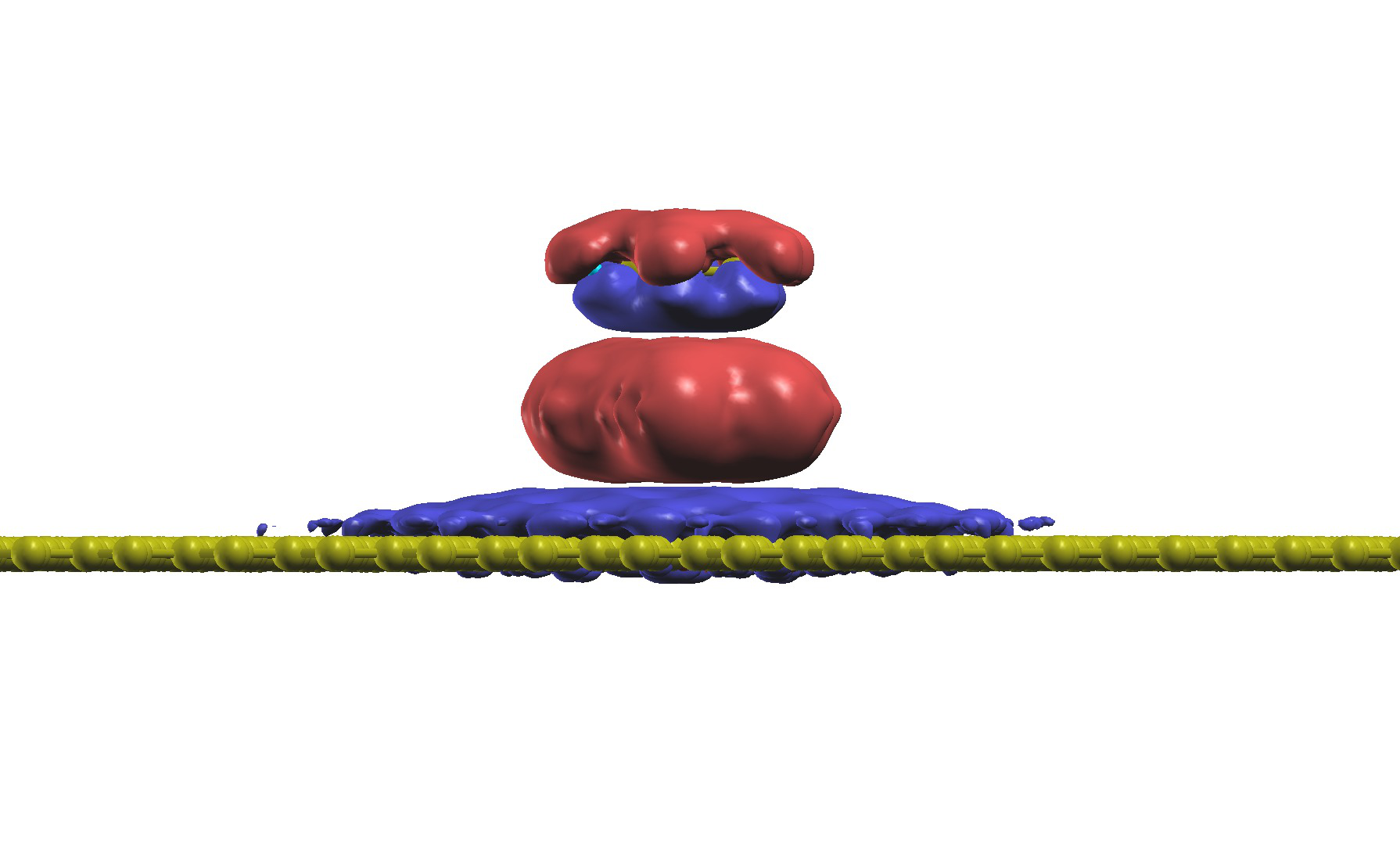}}%
}
\subfloat[]{\label{fig:f5_2}%
  \fbox{\includegraphics[width=0.188\textwidth]{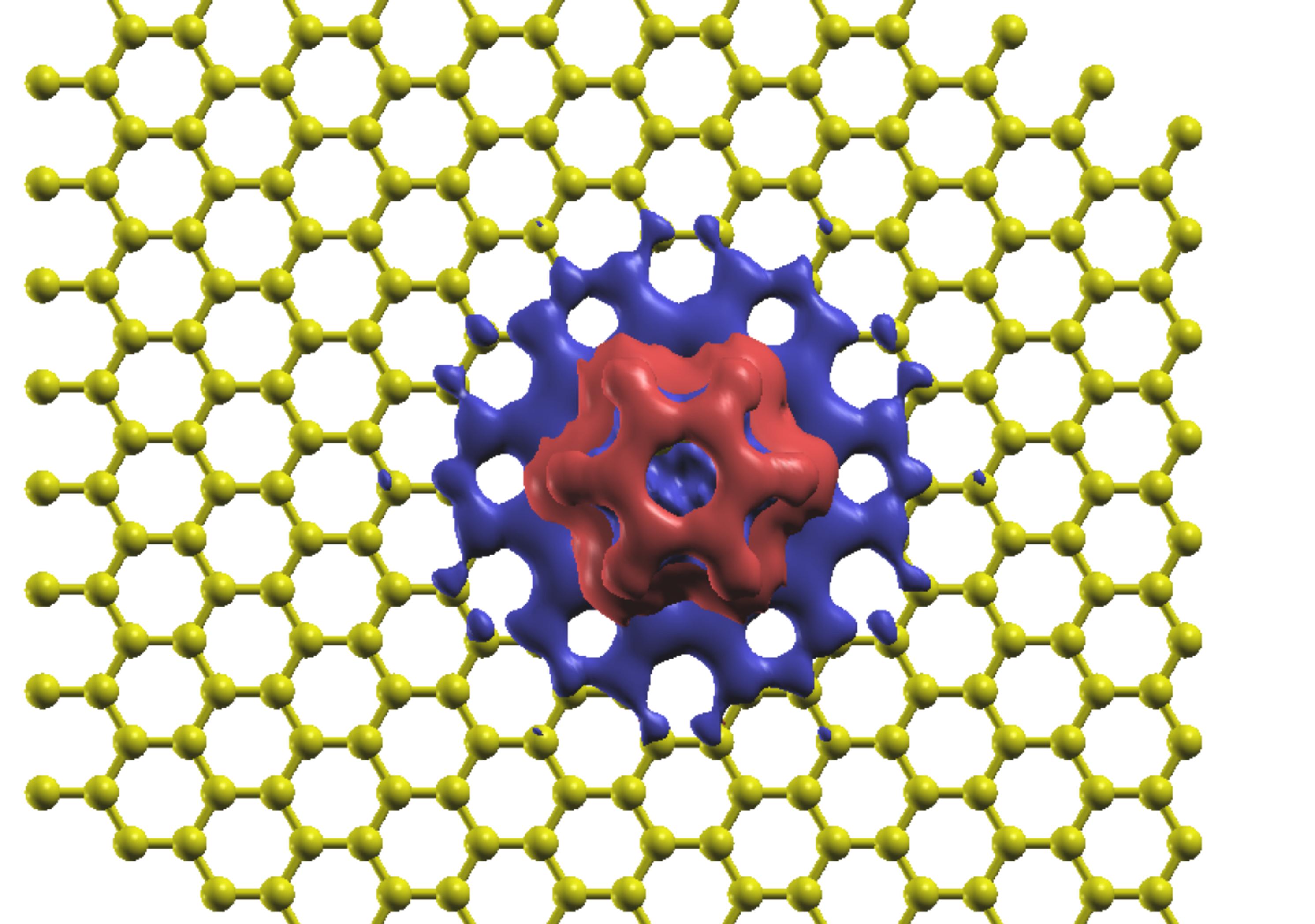}}%
}

\subfloat[]{\label{fig:f5_3}%
  \fbox{\includegraphics[width=0.22\textwidth]{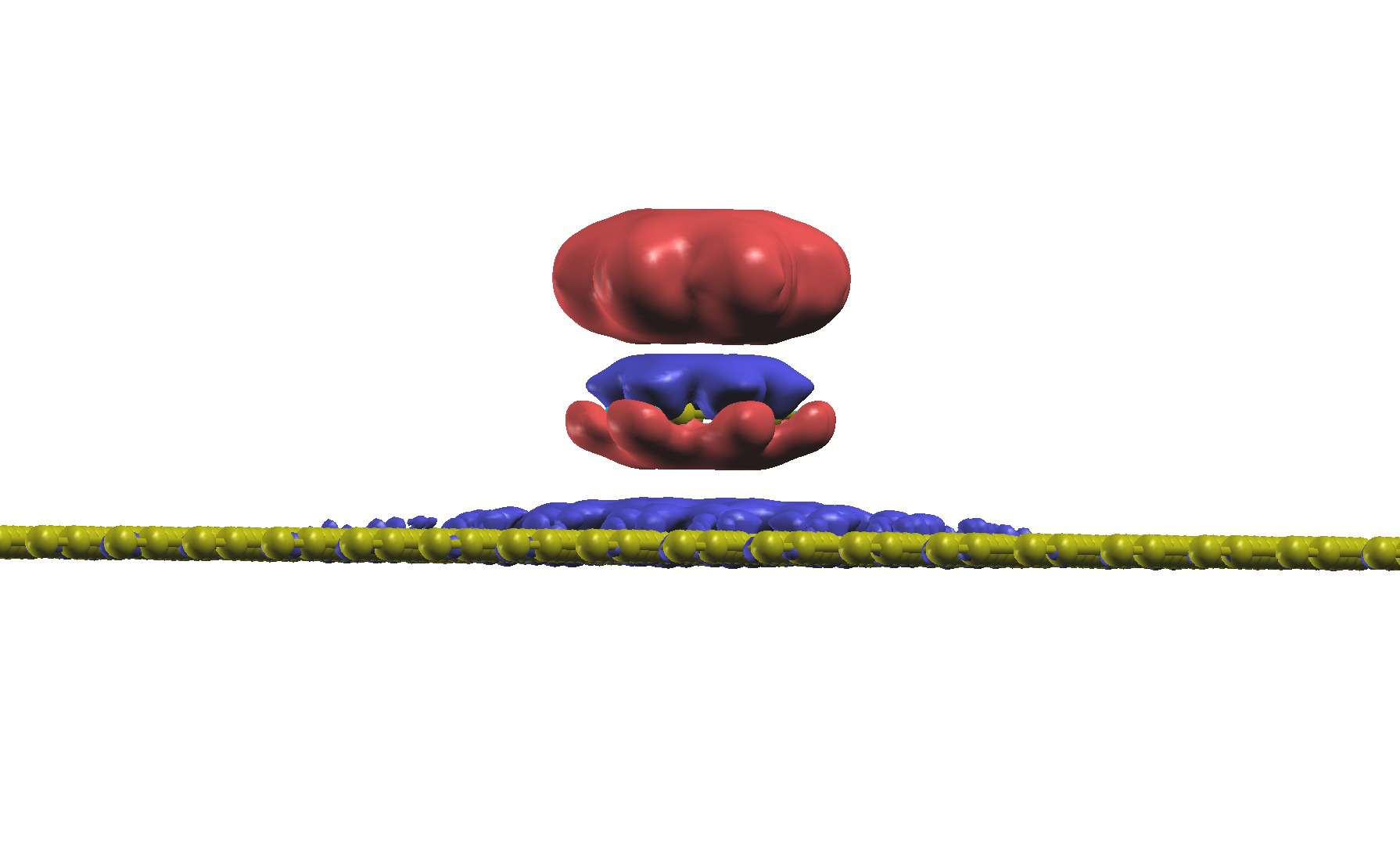}}%
}
\subfloat[]{\label{fig:f5_4}%
  \fbox{\includegraphics[width=0.192\textwidth]{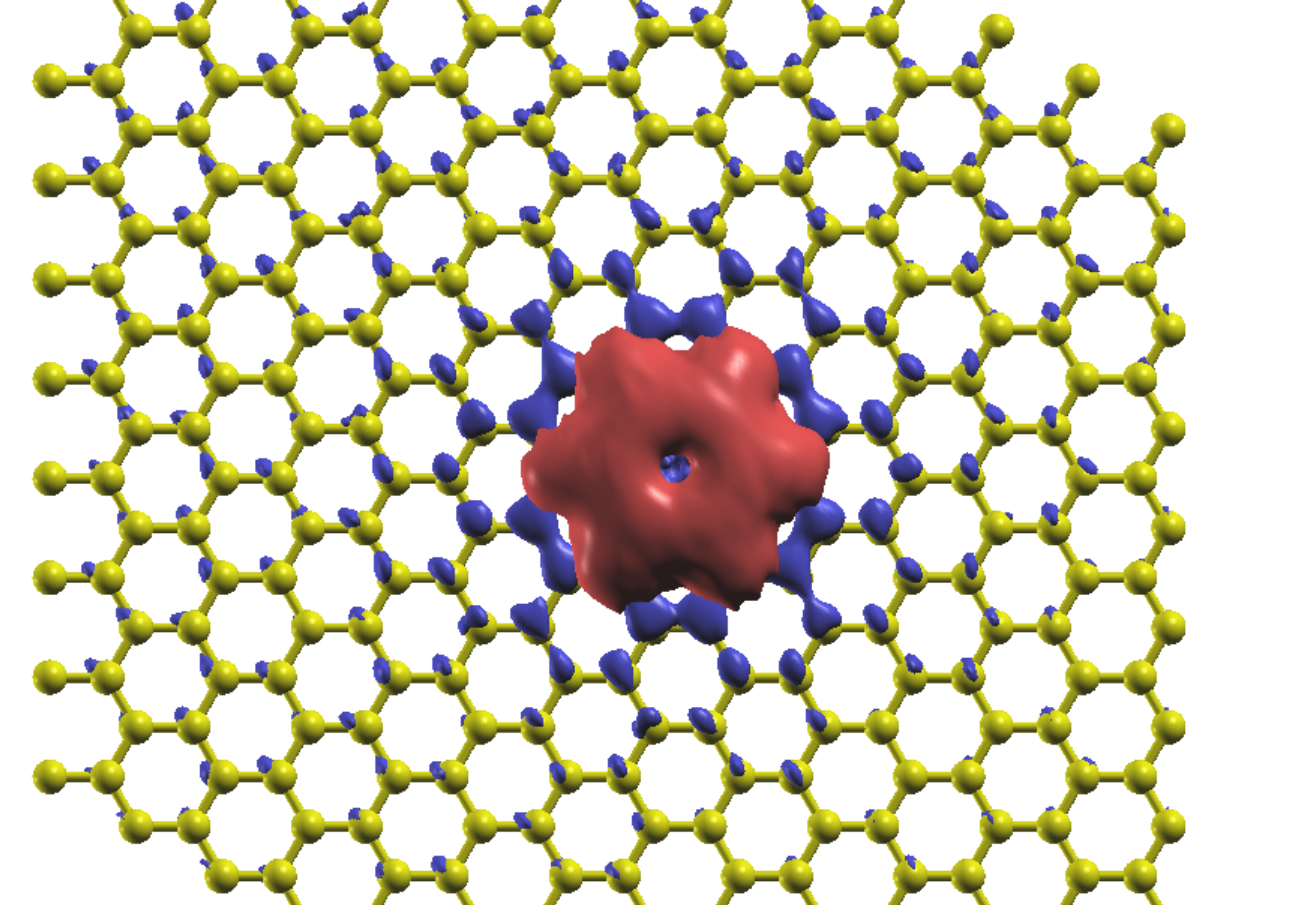}}%
}

\subfloat[]{\label{fig:f5_6}%
  \fbox{\includegraphics[width=0.40\textwidth]{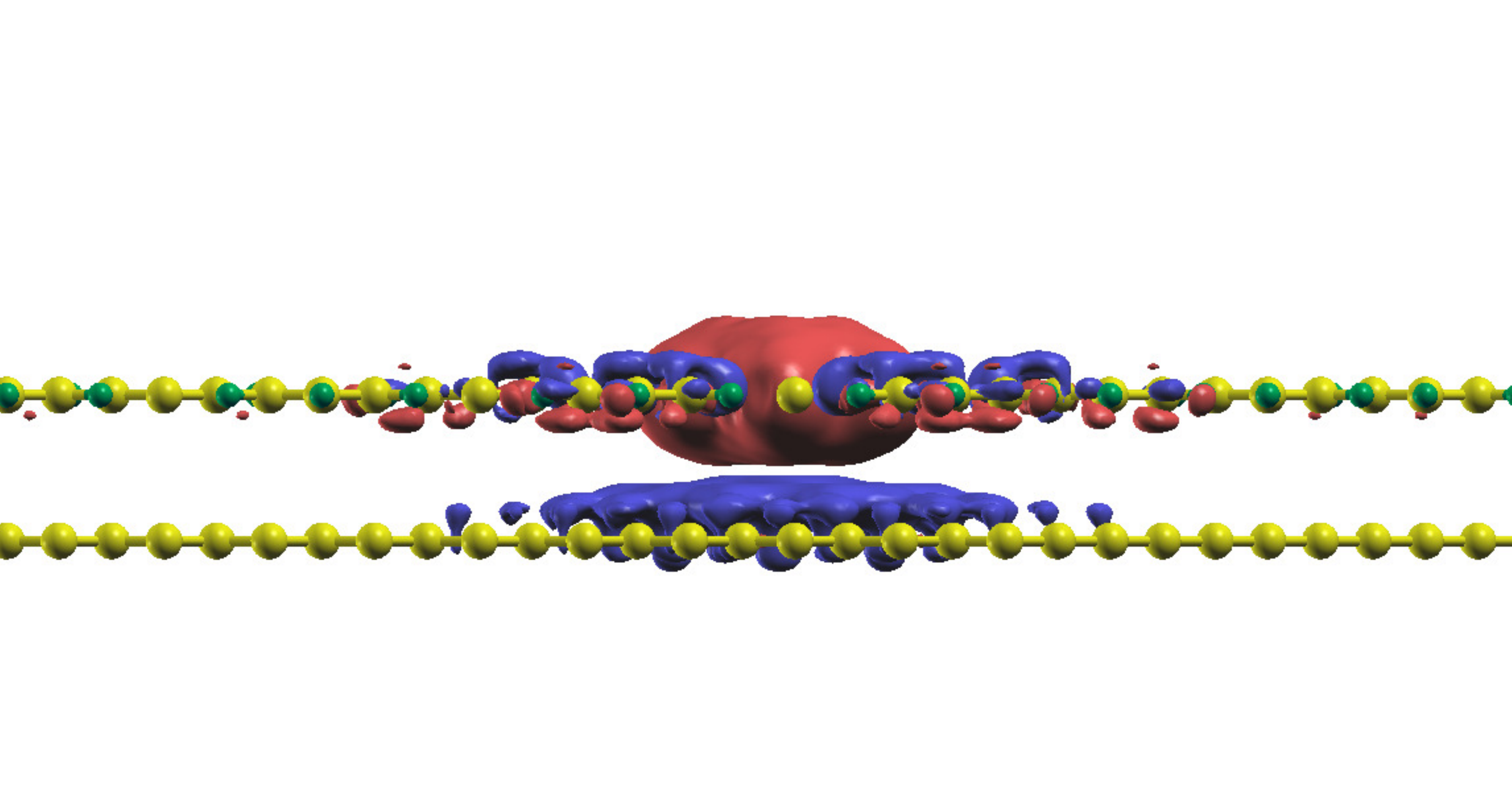}}%
}

\subfloat[]{\label{fig:f5_7}%
  \fbox{\includegraphics[width=0.40\textwidth]{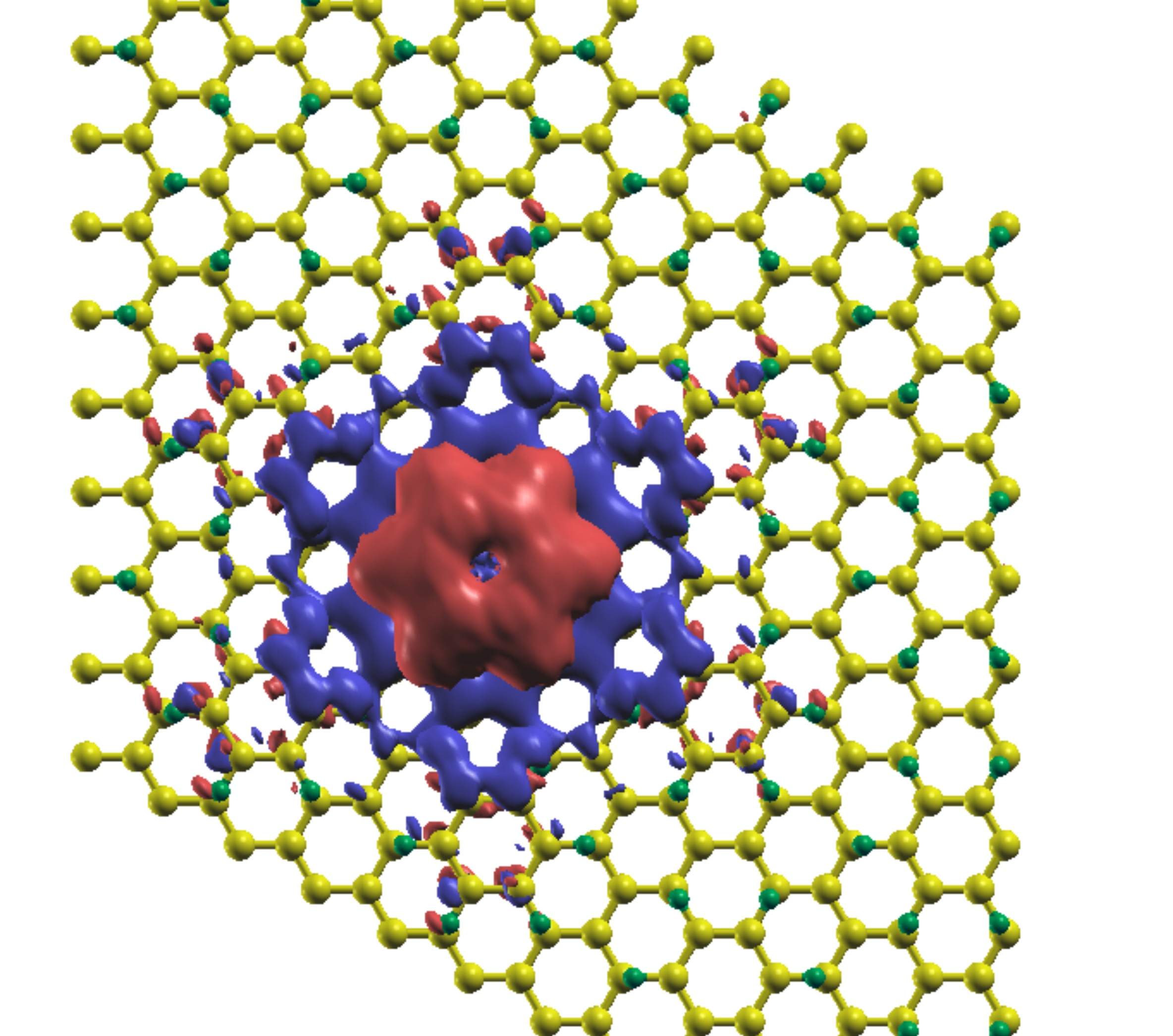}}%
}
\caption{Image charge formation for various configurations in which more than one benzene molecule is
present. Panels (a) and (b) respectively show the side and top view of the isovalue of $\Delta \rho(\mathbf{r})$ 
after transferring one electron from the molecule to the sheet for the $\rm{H_{M1}}$ case. Panels (c), (d) and 
(e), (f) show plots of the same quantity for the cases of $\rm{H_{M2}}$ and $\rm{H_L}$, respectively.}\label{fig:he}
\end{figure}

In real interfaces between organic molecules and a substrate, molecules usually are not found isolated 
but in proximity to others. It is then interesting to investigate the effects that the presence of other benzene 
molecules produce of the charge transfer energies of a given one. To this end we select three representative 
configurations. In the first one, $\rm{H_{M1}}$, the graphene sheet is decorated with two benzene molecules,  
one at 3.4~{\AA} while the other is placed above the first at 6.8~\AA\ from the graphene plane. We then 
calculate the charge transfer energies of the middle benzene (the one at 3.4~{\AA} from the sheet). The 
excess charge on different parts of the system (image charge), after transferring one electron to the 
sheet, is displayed in Fig.~\ref{fig:he}(a) and Fig.~\ref{fig:he}(b). The second 
configuration, $\rm{H_{M2}}$, is identical to the first one but now we calculate the charge transfer energies of the 
molecule, which is farther away from the graphene sheet, namely at a distance of 6.8~{\AA}. For this configuration, 
the excess charge after a similar charge transfer is shown in Fig.~\ref{fig:he}(c) and Fig.~\ref{fig:he}(d). In the third 
configuration, $\rm{H_L}$, we arrange multiple benzene molecules in the same plane. The molecules are in 
close proximity with each other although their atomic orbitals do not overlap. Charge transfer energies are 
then calculated with respect to one benzene molecule keeping the others neutral and an isovalue plot for similar 
charge transfer is shown in Fig.~\ref{fig:he}(e) and Fig.~\ref{fig:he}(f). 

The charge transfer energies calculated for these three configurations are shown in Tab.~\ref{table:configuration}. 
If one compares configurations where the molecule is kept at the same distance from the graphene plane,
such as the case of H($d=3.4$~\AA), $\rm{H_{M1}}$ and $\rm{H_L}$ or of H($d=6.8$~\AA) and $\rm{H_{M2}}$,
it appears clear that the presence of other molecules has some effect on the charge transfer energies. In particular 
we observe than when other molecules are present both $E_{\rm CT}^+$ and $E_{\rm CT}^-$ get more shallow,
i.e. their absolute values is reduced. Interestingly the relative reduction of $E_{\rm CT}^+$ and $E_{\rm CT}^-$
depends on the details of the positions of the other molecules (e.g. it is different for $\rm{H_{M1}}$ and $\rm{H_L}$)
but the resulting renormalization of the HOMO-LUMO gap is essentially identical [about 25~meV when going from
H($d=3.4$~\AA) to either $\rm{H_{M1}}$ or $\rm{H_L}$].

This behaviour can be explained in terms of a simple classical effect. Consider the case of $\rm{H_{M1}}$ for
example. When one transfers an electron from the middle benzene to the graphene sheet the second benzene
molecule, placed above the first, remains neutral but develops an induced charge dipole. The moment of such
dipole points away from the charged benzene and lowers the associated electrostatic potential. Importantly,
also the potential of graphene will be lowered. However, since the potential generated by an electrical dipole
is inversely proportional to the square of distance, the effect remains more pronounced at the site of the middle 
benzene than at that of the graphene sheet. A similar effect can be observed for an electron transfer from the 
graphene sheet to the middle benzene and for the $\rm{H_L}$ configuration. 

In the case of $\rm{H_{M2}}$, the system comprising the topmost benzene (from which we transfer charge) 
and the graphene plane can be thought of as a parallel-plate capacitor. The work, $W$, done to transfer a 
charge $Q$ from one plate to the other is $W={Q^2}/{2C}$ where $C$ is the capacitance, which in turn 
is proportional to the dielectric constant of the medium enclosed between the plates. Hence, at variance with
the case of H(d=6.8~{\AA}), the space in between the molecule and the graphene sheet is occupied by a molecule
with finite dielectric constant and not by vacuum. This results in a reduction of $W$, so that the charge transfer 
energies for $\rm{H_{M2}}$ are smaller than those for H(d=6.8~{\AA}).
 
\subsection*{Classical electrostatic approximation} 
\begin{figure}
\vspace{1cm}
 \centering
  \includegraphics[width=0.44\textwidth]{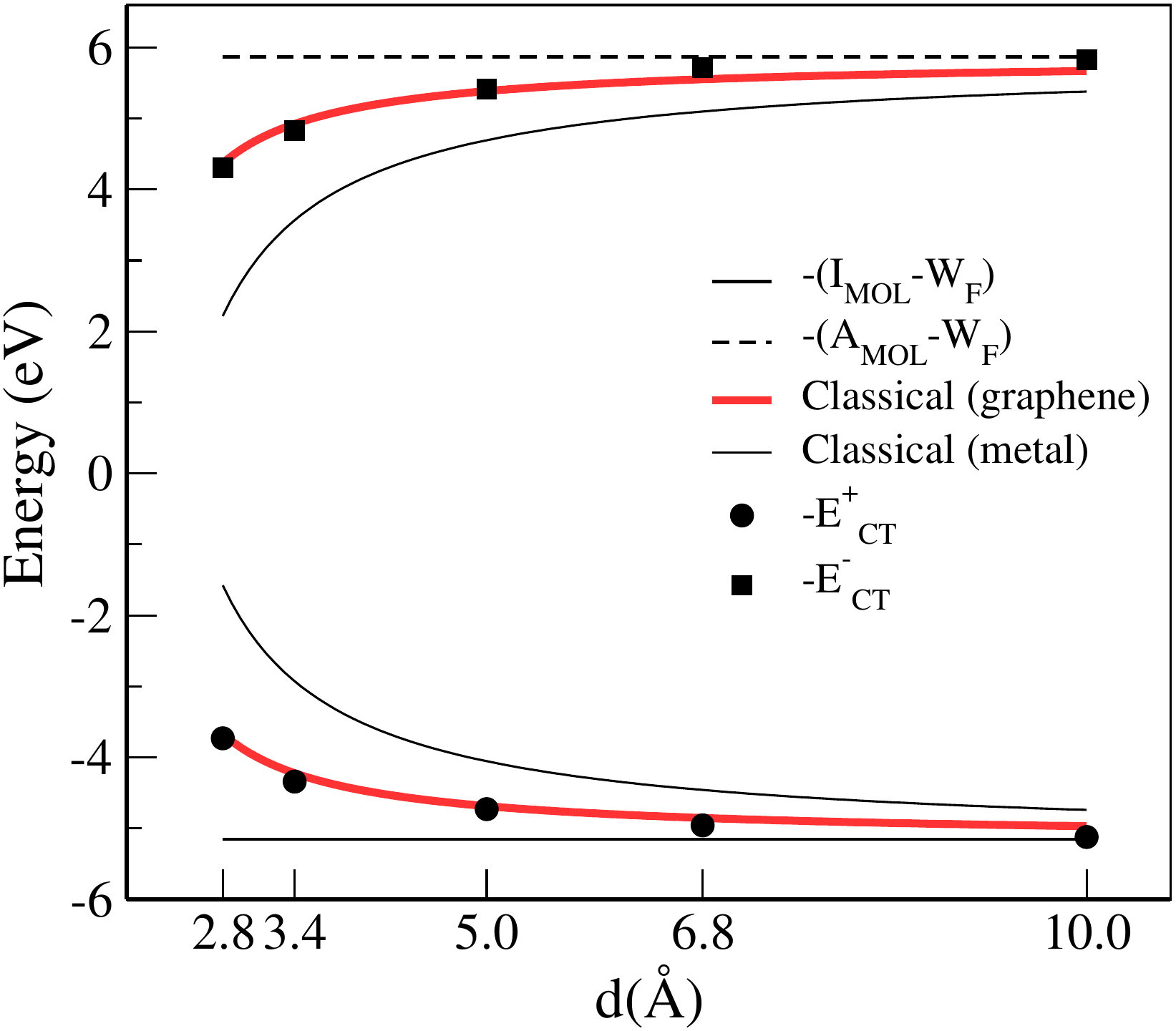}
  \captionsetup{justification=justified, singlelinecheck=false}
  \caption{$-E^+_{\rm CT}$ (circles) and $-E^-_{\rm CT}$ (squares) calculated for different 
  molecule-to-substrate distances. The CDFT results are seen to agree well with the classically 
  calculated curve given in red. The horizontal lines mark the same quantities for isolated an molecule
  (gas-phase quantities). The continuous black line shows the position of the classically calculated 
  level curve for adsorption on a perfect metal $\epsilon=\infty$.}
  \label{fig:classical_plot}
 \end{figure}

Finally, we show that our calculated energy levels alignment can be obtained from a classical electrostatic 
model. If one approximates the transferred electron as a point charge and the substrate where the image charge 
forms as an infinite sheet of relative permittivity $\epsilon$ then, for a completely planar distribution of the bound 
surface charge, the work done by the induced charge to take an electron from the position of the molecule 
(at a distance $d$) to infinity is 
\begin{equation}\label{work}
W=-\frac{1}{4\pi\epsilon_0}\left(\frac{\epsilon-1}{\epsilon+1}\right)\frac{q^2}{4d}\:.
\end{equation}
Hence, this electrostatic approximation predicts that the presence of the substrate lowers the ${\rm LUMO}$ of 
the molecule by $-\frac{1}{4\pi\epsilon_0}\left(\frac{\epsilon-1}{\epsilon+1}\right)\frac{q^2}{4d}$ with respect to 
the corresponding gas-phase value. However, the actual image charge is not strictly confined to a 2D plane 
but instead spills out over the graphene surface. We can account for such non-planar image charge distribution
by introducing a small modification to the above expression~\cite{Lang} and write the LUMO at a height $d$ as 
\begin{equation}\label{LUMOd0}
{\rm LUMO}(d)={\rm LUMO}(\infty)-\frac{1}{4\pi\epsilon_0}\left(\frac{\epsilon-1}{\epsilon+1}\right)\frac{q^2}{4(d-d_0)}\:, 
\end{equation}
where $d_0$ is the distance between the centre of mass of the image charge and the substrate plane and 
${\rm LUMO}(\infty)$ is the gas-phase LUMO (the electron affinity). A similar argument for the HOMO level 
shows an elevation of same magnitude due to the presence of the substrate. In Fig.~\ref{fig:classical_plot} we plot the 
charge transfer energies and show that they compare quite well with the curves predicted by the classical model
by using an effective dielectric constant of 2.4 for graphene~\cite{Wunsch}. 
When drawing the classical curves we have used an approximate value, $d_0=1.7$~\AA, which provides an excellent 
estimate for smaller distances, $d$. It is worth noting that for larger distances, though the actual value of $d_0$ 
should be much less, the overall effect of $d_0$ is very small and almost negligible. In the same graph, we have 
also plotted the classical curves corresponding to benzene on a perfectly metallic ($\epsilon=\infty$) surface. This 
shows that the level renormalization of benzene for physisorption on graphene is significantly different from that on 
a perfect metal, owing to the different screening properties of graphene.

\section*{Conclusion}
We have used CDFT as implemented in the SIESTA code to calculate the energy levels alignment of 
a benzene molecule adsorbed on a graphene sheet. In general the charge transfer energies depend on 
the distance between the molecule and the graphene sheet, and this is a consequence of the image 
charge formation. Such an effect cannot be described by standard Kohn-Sham DFT, but it is well captured
by CDFT, which translates a quasi-particle problem into an energy differences one. With CDFT we have 
simulated the energy level renormalization as a function of the molecule-to-graphene distance. These
agree well with experimental data for an infinite separation, where the charge transfer energies coincide
with the ionization potential and the electron affinity. Furthermore, an excellent agreement is also obtained 
with $GW$ calculations at typical bonding distances. Since CDFT is computationally inexpensive we have
been able to study the effects arising from bonding the molecule to a graphene structural defect and
from the presence of other benzene molecules. We have found that a Stone-Wales defect does not affect 
the energy level alignment since its electronic density of state has little amplitude at the graphene Fermi level.
In contrast the charge transfer energies change when more then a molecule is present. All our results can
be easily rationalized by a simple classical electrostatic model describing the interaction of a point-like charge
and a uniform planar charge distribution. This, at variance to the case of a perfect metal, takes into account
the finite dielectric constant of graphene.

\section*{Acknowledgment}
This work is supported by the European Research Council, Quest project. Computational resources have been 
provided by the supercomputer facilities at the Trinity Center for High Performance Computing (TCHPC) and at 
the Irish Center for High End Computing (ICHEC). Additionally, the authors would like to thank Dr. Ivan Rungger and Dr. A. M. Souza for helpful discussions.

\end{document}